\documentclass[review]{elsarticle}
\usepackage{hyperref}

\journal{Journal of \LaTeX\ Templates}

\usepackage{color}
\usepackage{amsmath}
\usepackage{graphicx}

\biboptions{authoryear}

\begin{document}


\begin{frontmatter}

\title{Interferometric calibration and the first elevation observations at EKB ISTP SB RAS
radar at 10-12 MHz}

\author{Oleg I.Berngardt}
\address{Lermontova Str., 126A, Irkutsk, Russia, ISTP SB RAS}
\ead{berng@iszf.irk.ru}
\author{Roman R.Fedorov}
\address{Lermontova Str., 126A, Irkutsk, Russia, ISTP SB RAS}
\ead{fedrr@iszf.irk.ru}
\author{Pavlo Ponomarenko}
\address{Physics building, 116 Science Place,Saskatoon, SK, S7N 5E2 , Canada, University of Saskatchewan}
\ead{pbp672@usask.ca}
\author{Konstantine V.Grkovich}
\address{Lermontova Str., 126A, Irkutsk, Russia, ISTP SB RAS}
\ead{grkovich@iszf.irk.ru}






\begin{abstract}
The method for calibrating elevation measurements at EKB ISTP SB RAS
radar obtained for the period 20/09/2019 - 18/11/2019 is presented. 
The calibration method is a modernization of the method for calibrating
radar by meteor trails.
The main difference of the method is the use of not a statistically
processed FitACF data, but the full waveform of the signals scattered on
the meteor trails. Using the full waveform makes it possible to more
reliably distinguish meteor scattering from other possible scattered
signal sources, and to determine meteor heights from the trail lifetime
using the NRLMSIS-00 model. A comparison of the results
with the results of E-layer calibration method shows a good agreement.
The first examples of regular elevation observations
at the EKB ISTP SB RAS radar are presented, and their preliminary
interpretation is given.
\end{abstract}


\end{frontmatter}


\section{Introduction}

The coherent decameter radars of Super Dual Auroral Network (SuperDARN)
\citep{Greenwald_1995,Chisham_2007,Nishitani_2019} and similar radars
\citep{Berngardt_2015} are effective instruments
for studying the magnetosphere-ionosphere interaction and for the
upper atmosphere monitoring by radiowave scattering technique. Radars are the instruments with the regular
spatial resolution 15-45km, the regular temporal resolution 1-2 minutes,
the maximum range 3500-4500km, and nearly $50^{o}$ field-of-view in azimuth.

The decameter radars operate at 8-20 MHz, so ionospheric
refraction plays a significant role.
The sources of the scattered signals received by the radars
are: the meteor trail scattering in D- and E- layers of the ionosphere \citep{Yukimatu_2002},
scattering from E- and F-layer ionospheric irregularities elongated
with the Earth magnetic field \citep{Greenwald_1995,Chisham_2007},
signals scattered by surface (ground and water) irregularities and 
detected due to refraction of radiowaves in the ionosphere \citep{Ponomarenko_2010,Greenwood_2011},
mesospheric echoes \citep{Hosokawa_2004,Hosokawa_2005,Ogunjobi_2015}
and other possible mechanisms \citep{Ponomarenko_2016}. The propagation
trajectory of the radio wave affects both the accuracy of identifying
the scattered signal type\citep{Bland_2014b} and the accuracy of determining
the parameters of ionospheric irregularities: their velocity \citep{Ponomarenko_2009,Gillies_2011},
altitude \citep{Koustov_2007,Ponomarenko_2009} and geographical coordinates
\citep{Villain_1984,Ponomarenko_2009,Berngardt_2015b}. Therefore,
the trajectory estimation is a complex theoretical and computational
problem and is the basis for solving important practical problems
of using decameter coherent radars for ionospheric and magnetospheric monitoring.

One of the main methods to improve the accuracy of the radiowave propagation trajectory
estimate are elevation (vertical angle of arrival, interferometric) observations
\citep{Villain_1984}. Due to the relatively long wavelength, comparable
with the distance from the antenna to the ground surface, the thermal
variations in radio cables and the analog parts of the transmitter and
receiver parts, the elevation observations require a regular phase
calibration \citep{Chisham_2013,Chisham_2018,Ponomarenko_2018}. Due to the 
potential dynamics of the calibration parameters, such
a technique should be automatic. 

The main problem with calibration of SuperDARN interferometry is that 
for direct calibration using a fixed source at a given location
one needs to put a target at about 100 km altitude and about 300-500~km 
range due to the scale of measurements. It is practically 
impossible and other techniques are used. 
Currently at SuperDARN radars the following basic calibration methods are used: 
by using the position of the ground scatter signal\citep{Ponomarenko_2015}; by
using the signals scattered in the E-layer of the ionosphere \citep{Ponomarenko_2018};
by using the signals scattered by the meteor trails\citep{Chisham_2013,Chisham_2018}; and 
by using the ionospheric targets with known geographic location, for example heater-induced 
artificial irregularities \citep{Burrell_2016}.

Ground scatter calibration is based on adjusting the phase difference
between the main and interferometer antenna arrays in such a way that the
resulting elevation approaches the theoretically expected zero values at the
far edge of the ground scatter band for any given 'hop' \citep{Ponomarenko_2015}. 
The main problem of the method is the intrinsically high dynamics in
the ground scatter range distribution due to strong variability of the
ionospheric parameters, which requires visual analysis of the data and
effectively prohibits automatic calibration process.

Less affected by the ionospheric refraction and therefore more suitable for automatic
processing are the methods based on using scattering in E- and D- layers - 
the lower part of the ionosphere.

In the case of using the signals scattered in the E-layer
of the ionosphere, the calibration procedure is based on matching
the observed phase distribution from the farthest ranges of the E-layer
echoes, where the elevation angle is expected to be nearly zero, with
the simulated distribution produced by using a simple statistical
model of the E-layer backscatter returns \citep{Ponomarenko_2018}.
Using this approach allows one to improve the accuracy of calibration
in comparison with that in \citep{Ponomarenko_2015} due to much lower 
variability in the E-layer altitude as compared to that of the F-layer. 
Furthermore, this approach does not require taking into account a refraction 
in the ionosphere which
allows creating an automatic calibration algorithm. In this case,
the calibration is performed over the data produced by standard
SuperDARN programs: the phase of the correlation function between
the signals received at main and interferometric
antenna arrays and averaged over the number of soundings. 
Another technique presented in \citep{Chisham_2013,Chisham_2018} allows calibrating measurements by adjusting the phase offset in such a way that the effective altitude of scattering at meteor trails (D- and E-layers) is the same across several range gates.

The problem of the both methods is the complexity of independent measurement
of the height of the scattering irregularity, found either by optimization
of the residual functional \citep{Chisham_2018}, by measuring average spectral widths \citep{Chisham_2013},
or by substituting a model height \citep{Ponomarenko_2018}. Another
problem of these methods is detecting scattered
signals of necessary kind and differing them from other signals, because misidentification
leads to processing errors. Using large amount of sounding data,
however, can mitigate these problems
\citep{Chisham_2018,Ponomarenko_2018}.

In this paper, we present a method for calibrating the elevation measurements
at EKB ISTP SB RAS radar by using the signals scattered on meteor
trails, improving \citep{Chisham_2013,Chisham_2018} technique. The problems of determining the scattering height and detecting
the scattering type are solved by independent algorithms. They are based
on the physical mechanisms of the meteor echo formation and on the automatic analysis of the received signals quadrature components. 

\section{Calibration method}

\begin{figure}
\includegraphics[scale=0.25]{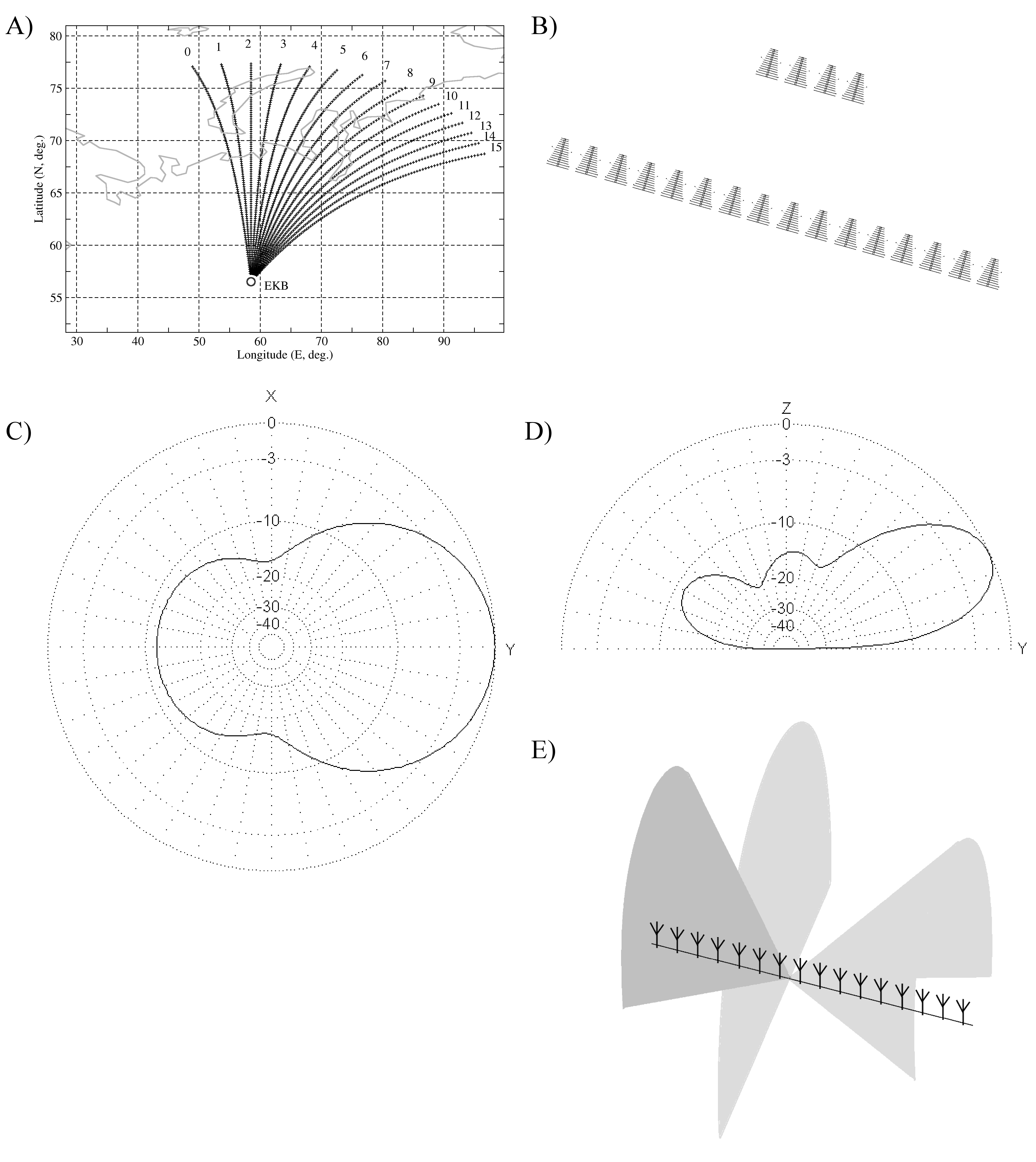}
\caption{A) EKB ISTB SB RAS radar location and field of view; 
B) Antenna field geometry; C-D) DLP11
antenna pattern at 10MHz in the horizontal and vertical plane correspondingly
(model calculations).
E) Antenna pattern
cone for 3 different azimuths; 
}
\label{fig:fig1}
\end{figure}

The EKB ISTP SB RAS \citep{Berngardt_2015} is mid-latitude coherent
radar installed in 2012 by the Institute of Solar-Terrestrial Physics,
Siberian Branch of the Russian Academy of Sciences (ISTP SB RAS) in
the Arti Observatory of Institute of Geophysics, Ural Branch of the
Russian Academy of Sciences ($56.43^{o}N,\,58.56^{o}E$, Sverdlovsk
Region, Russia, 180 km south-west from Ekaterinburg). The radar electronics
was produced for ISTP by University of Leicester (UK) and is an analogue
of the SuperDARN CUTLASS stereo radar\citep{Lester_2004}. The radar operates 
at 8-20MHz and 
uses standard SuperDARN radar software to process sounding data. 
The antenna pattern of the radar
is formed by 20 DLP11 (Titanex Gmbh, Germany) log-periodic
antennas located in two linear equidistant phased arrays: transmit-receive one
(16 antennas, main array) and receive one (4 antennas, interferometric array). This construction of the
radar allows one to scan the $52^{o}$ radar field-of-view in 16 fixed
directions (beams). The radar field of view and beam positions are
shown in Fig.\ref{fig:fig1}A. The azimuthal resolution of the radar
is $3-6^{o}$, depending on the frequency. For elevation angle observations
the combination of main and interferometric arrays is used. The antenna field
geometry is shown in Fig.\ref{fig:fig1}B. The signals received from
each antenna in the phased array are summarized with necessary
phase differences for the radar beam forming. The DLP11 antenna
pattern is quite complex, frequency dependent and has the significant
back lobe. The model calculations of the antenna pattern using MMANA-GAL
software \citep{MMANAsite} are shown in Fig.\ref{fig:fig1}C-D. Such
an antenna pattern leads to a noticeable number of the signals received
in the back lobe of the antenna pattern and should be taken into account
when processing the data. 

The measurements of the phase difference between the signals received
by main and interferometric antenna arrays provide the elevation
angles estimations and require calibration. Due to the linear orientation
of the main and interferometric arrays, the antenna pattern at a fixed
azimuth for each of the arrays corresponds to the surface of a cone
(Fig.\ref{fig:fig1}E), and the azimuth should be taken into account
when calculating elevation. The $2\pi n$ uncertainty in the calculation of
the phase caused by the large distance between main and interferometric
arrays should be taken into account too.

To calibrate elevation observations we choose the meteor calibration
method \citep{Chisham_2013,Chisham_2018}, modified to use wide capabilities
of the EKB ISTP SB RAS radar to process meteor trail scattering. The
calibration method \citep{Chisham_2018} uses processed data, extracts
meteor signals by range (\textless{}400km) and determines the average
meteor trail height using algorithmic optimization. In \citep{Chisham_2013} 
the meteor trail height is determined from average spectral width.

Our method uses exclusively
the waveform of signals scattered from meteor trails measured on two
phased arrays. The scattering height is calculated based on the shape
of the scattered signal using the reference model of the neutral atmosphere
NRLMSIS-00 \citep{NRLMSIS_2002}.

There are two types of scattering on meteor trails, differing by the
ratio of the trail plasma frequency to the sounding frequency: underdense
and overdense echo.

To calibrate the
elevation observations we use underdense echo because their
cross-section exponentially decrease with time and they are easy to detect.
Their dynamics
is controlled by recombination processes and in the first approximation
is related to the diffusion coefficient at the burn height \citep{Jones_and_Jones1990JATP...52..185J} 
so their altitude can also be detected from the radar data.
At EKB radar we use the algorithm for detecting underdense meteor
trail echoes, similar to the detection algorithms at specialized meteor
radars and SuperDARN radars \citep{Tsutsumi_BPMR_autodetect1999,Tsutsumi_SDARN_meteorwindIQ}
but adjusted to use EKB radar features.

\begin{figure}
\includegraphics[scale=0.4]{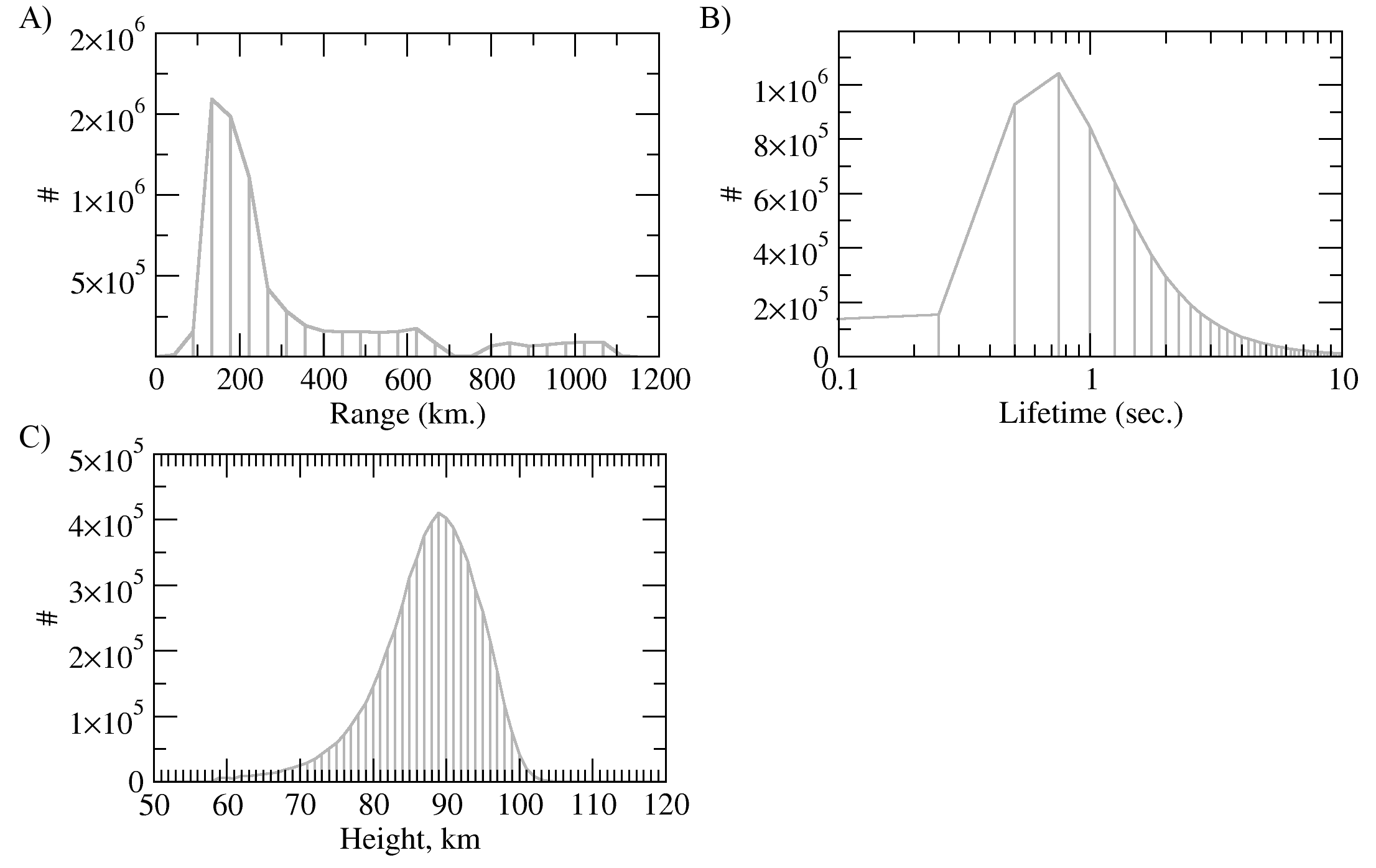}
\caption{Distributions of meteors (histograms) observed on the EKB radar during
2017-2019 by ranges (A), lifetimes (B) and burn heights (C).}
\label{fig:fig2}
\end{figure}

Search and selection of meteor trail scattering at EKB radar is carried
out using the following scheme:
\begin{enumerate}
\item Search for bursts of signal level spatially localized by range and
azimuth;
\item Fit separately the amplitude and the phase of the signal by
the model
\begin{equation}
\left\{ \begin{array}{l}
A_{m}(t)=\theta(t)A_{0}e^{-\frac{t}{\tau}}\\
\phi_{m}(t)=\theta(t)(st+\phi_{0})
\end{array}\right.\label{eq:meteor_model}
\end{equation}
where $A_{m}(t)$ is the model amplitude; $\phi_{m(t)}$ is the model
phase; $A_{0}$ and $\phi_{0}$ are the initial amplitude and phase;
$s$ is linear phase change coefficient proportional to the Doppler
drift velocity, $\tau$ is characteristic trail lifetime.
\item In the case of high fitting accuracy, interpret the signal as
meteor trail scattering with corresponding lifetime and velocity.
The fitting accuracy is determined by the variance of the amplitude
and phase of the received signal relative to their model values (\ref{eq:meteor_model}).

\end{enumerate}
In this paper we use the
characteristic lifetime, associated
with the diffusion coefficient at the burning height. 
The trail lifetime distribution is shown in Fig.\ref{fig:fig2}B.
Using the neutral
atmosphere model and  trail lifetime
makes it possible to determine the coordinates of
the scattering point without using elevation
observations. 

The characteristic meteor trail lifetime $\tau$ of the underdense
echo is defined by the diffusion coefficient $D$ at the burn height
\citep{Tsutsumi_SDARN_meteorwindIQ, Chisham_2013}:

\begin{equation}
\tau=\frac{\lambda^{2}}{32\pi^{2}D}\label{eq:meteor_lifetime}
\end{equation}
and the diffusion coefficient is: 

\begin{equation}
D=\frac{6.39*10^{-2}KT^{2}}{p}\label{eq:diffusion}
\end{equation}
where $\lambda$ is sounding signal wavelength; $T$ is absolute temperature;
$p$ is the pressure; $K$ is the mobility coefficient of the ions
in the meteor trail, usually taken as $2.2\cdot10^{-4}m^{2}s^{-1}V^{-1}$\citep{Tsutsumi_SDARN_meteorwindIQ}.

The burn height is determined by the NRLMSIS-00 model \citep{NRLMSISsite}
for a given time and coordinates, by iterative search over the heights.
The distribution of the burn heights calculated according to the EKB
radar data over the period 2017-2019 is shown in Fig.\ref{fig:fig2}C.
As can be seen from the figure, the main part of the distribution
corresponds to heights 80-100 km with the most probable burn height
89 km, which is in a good agreement with the results obtained earlier
\citep{7770504}.

Refraction at altitudes lower than the meteor's burn
altitudes is weak and the sounding signal trajectory is linear. Using
this approach the interferometer can be calibrated using the radar observations
only. 

\begin{figure}
\includegraphics[scale=0.45]{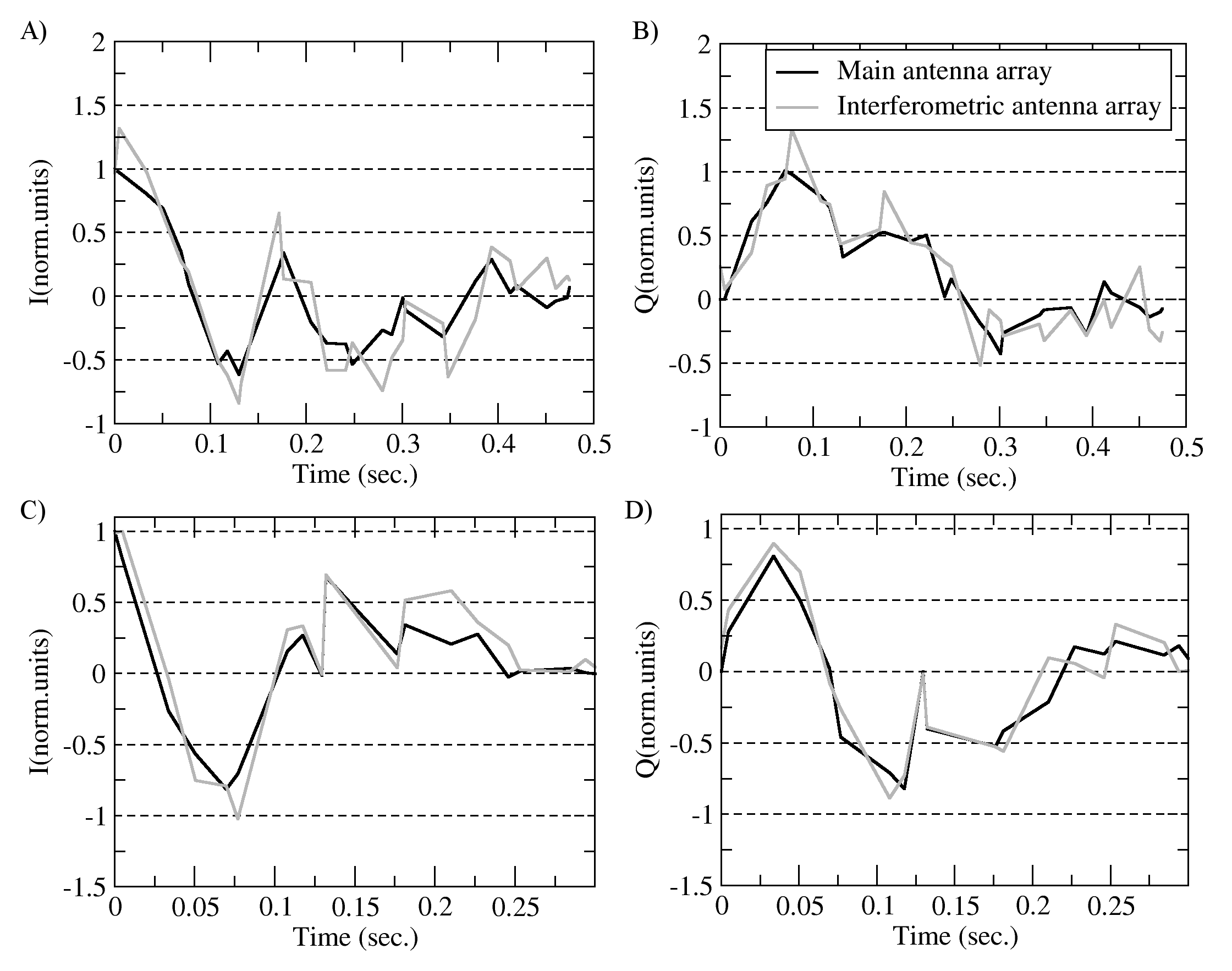}
\caption{Examples of meteor echoes recorded by different arrays - the black
line corresponds to the main array, the green - to the interferometric
array. On the left are the I-components of the received signals, on
the right are the Q-components. The amplitudes are normalized.}
\label{fig:fig3}
\end{figure}

Fig.\ref{fig:fig3}A-D shows examples of signals scattered on meteor
trails and received by the two arrays of the radar — the main (black
line) and intererometric (gray line), obtained with a high correlation
coefficient (R\textgreater{} 0.8). The figure demonstrates a good
correlation between the shapes of the real (left) and imaginary (right)
components of the received signal. A high degree of correlation of
the shapes of the received signal, its long lifetime (of the order
of 0.3-0.5 seconds) and its complex shape allow to detect meteors
and stably measure the phase difference between signals received by
the two antenna arrays with a good accuracy.

\begin{figure}
\includegraphics[scale=0.5]{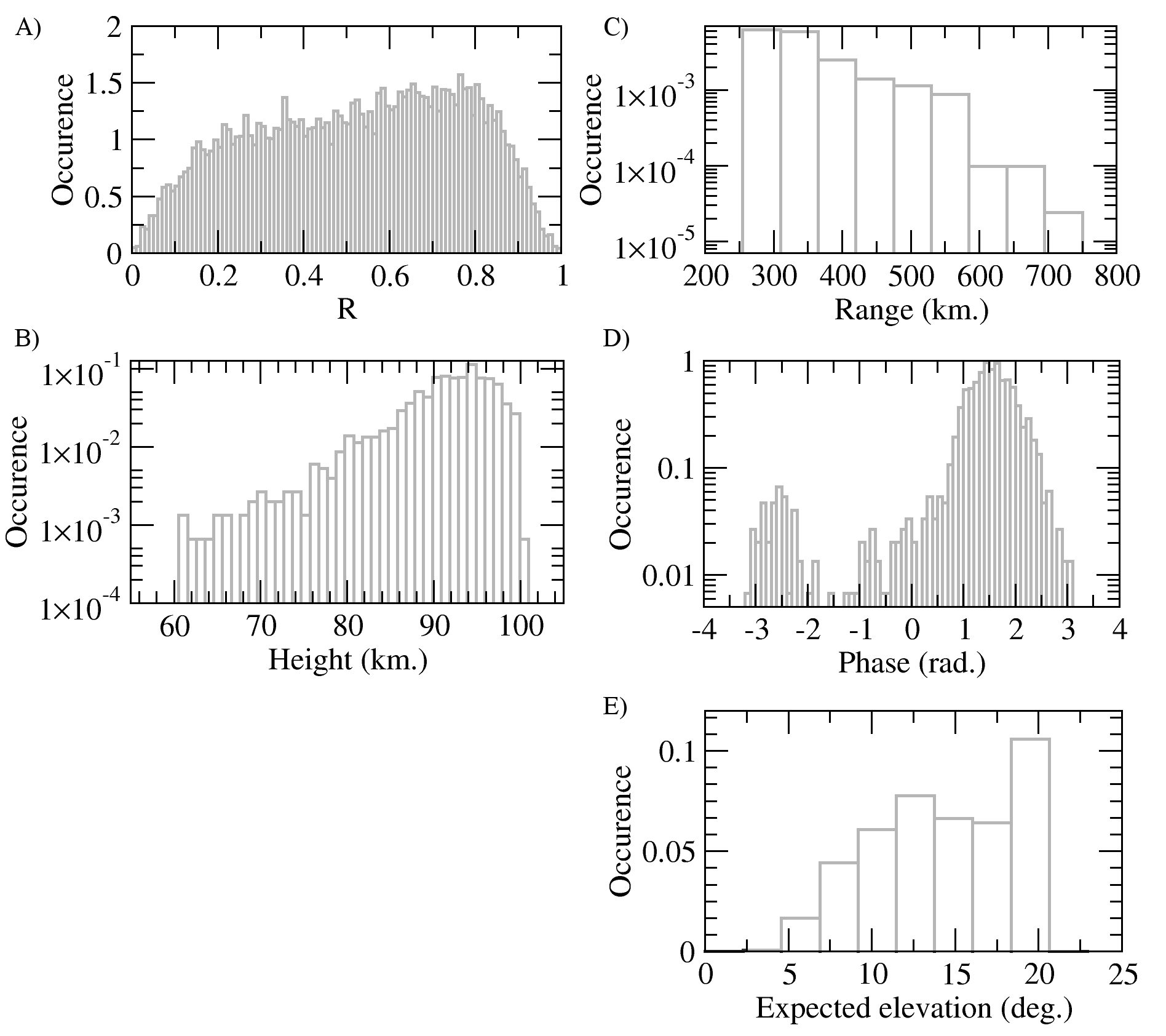}
\caption{Statistical distributions of the meteor trail characteristics used
for the calibration. A) is the distribution of the meteor correlation
coefficients between the antenna arrays; B) is the distribution of
filtered (250-750km, R\textgreater{} 0.8) meteors over altitude; C)
is the distribution of filtered (250-750km, R\textgreater{} 0.8) meteors
over the ranges; D) is the distribution of filtered (250-750km, R\textgreater{}
0.8) meteors over the estimated phase difference; E) is the distribution
of filtered (250-750km, R\textgreater{} 0.8) meteors over the expected
elevation angles.}
\label{fig:fig4}
\end{figure}

Fig.\ref{fig:fig4}A shows the distribution of the correlation coefficients
between signals received by different arrays. It can be seen from
the figure that the most probable value of the correlation coefficient
is 0.76, and the number of observations with a correlation coefficient
\textgreater{}0.8 is about 15\%. The high average number of observed
meteors (\textgreater{}200 per hour) and the large number of meteor
trail signals observed by both arrays allow using a large number of
meteor observations for EKB radar calibration. For the calibration,
we select only meteor data during the period 20/09/2019-18/11/2019
observed by both phased arrays at ranges above 250km and below 750km with 
a high correlation coefficient (R\textgreater{}0.8).
The restriction of ranges below 250km is used to reduce the influence
of spatial resolution to the accuracy of determining the elevation
angle and thereby to increase the accuracy of calibrated elevation
data. The upper limit 750km is used for additional decrease of the
number of possible ground scatter signals and signals scattered by
ionospheric irregularities on the resulting observation statistics.
The lower limit of the correlation coefficient is used to increase
the accuracy of determining the interference phase of the received
signals.

The total number of meteor trails used for calibration is about 1500.
The distributions of the meteor trails over the heights and the ranges
are shown in Fig.\ref{fig:fig4}B-C correspondingly. 

To calibrate the phase difference between the phased arrays, the model
and experimental phases have been compared. The experimental phase difference
was determined as the phase providing maximal cross-correlation coefficient
between the signals received by
both arrays.

The model phase difference is obtained from geometric considerations as the following:

\begin{equation}
\begin{array}{l}
\Delta\varphi_{mod,i}\approx2\pi\frac{\Delta_{y}cos\left(\theta_{i}\right)}{\lambda_{0,i}}cos\left(\alpha_{i}-atan\left(\frac{\Delta_{z}}{\Delta_{y}cos(\theta_{i})}\right)\right)\\
\alpha_{i}=asin\left\{ \frac{\left(R_{E}+H_{i}\right)^{2}-\left(R_{E}^{2}+R_{i}^{2}\right)}{2R_{E}R_{i}}\right\} 
\end{array}\label{eq:phase_model}
\end{equation}
where $\Delta_{y},\Delta_{z}$ are the displacement of the center
of the interference array relative to the center of the main array
in the horizontal and vertical directions correspondingly (56.4 and
-3.8 meters); $H_{i},R_{i}$ - are the height of the observation of
the i-th meteor trail and the distance to it, correspondingly; $\lambda_{0,i},\theta_{i}$
are the wavelength of the sounding signal and the ground azimuth
to the meteor trail relative to the center of the radar field-of-view,
correspondingly; $R_{E}$ is the radius of the Earth; $\alpha_{i}$
is the estimated elevation angle of the observed meteor trail.

It is traditionally supposed that the phase discrepancy between model and experimental phase observations is associated
with a total phase shift between the receivers and the difference
in the electric propagation lengths of the signal from the phased
arrays to the receivers \citep{Chisham_2018,Ponomarenko_2018}. As
a preliminary analysis showed, the EKB has a significant
non-linear component of the phase difference caused by the
hardware. Therefore, the calibration problem in this paper was solved
for the basic operating frequencies of the radar 10-12MHz only. This allows
us to neglect the frequency dependence of the phase difference and
reduce the calibration problem to the search for a single unknown parameter
$A$. The parameter should minimize the difference between the model
phase difference $\Delta\varphi_{mod,i}$ and the experimental 
phase difference
$\Delta\varphi_{exp,i}$ for all meteor trails:

\begin{equation}
\sum_{i=1}^{N}\left[\Delta\varphi_{exp,i}-(\Delta\varphi_{mod,i}+A)\right]^{2}=0\label{eq:fit_eq}
\end{equation}
here the summation is made over all the $N$ detected meteor
trails.

The distributions of the expected antenna phase difference and elevations
of the observed meteors are shown in Fig.\ref{fig:fig4}D-E. It can
be seen from Fig.\ref{fig:fig4}D that about 4\% of meteor tails are
observed in the back lobe of the antenna pattern (which corresponds
to phases \textless{}-2 radians), which should be taken into account
in the calibration algorithm. A small part of such trails allows us simply to 
remove them from consideration, and this should
not significantly affect the accuracy of the resulting algorithm.
The final algorithm for determining the calibration coefficient $A$
becomes a three-stage one:
\begin{enumerate}
\item based on the data set that simultaneously satisfies the conditions $R>0.8,R_{i}>250km,R_{i}<750km$,
the first approximation of the calibration coefficient $A_{0}$ is
made using (\ref{eq:fit_eq});
\item the observations in the back lobe of the antenna pattern are removed.
They are defined as observations that differ from the dependence found
at the first stage by more than 1 radian: $\left|\Delta\varphi_{exp,i}-(\Delta\varphi_{mod,i}+A_0)\right|>1$
;
\item based on the remaining data set, more accurate calibration coefficient
$A$ is found using (\ref{eq:fit_eq}).
\end{enumerate}
It should be noted that in contrast to the methods \citep{Chisham_2018,Ponomarenko_2018},
the main parameter used for calibration is not a statistical dependence
of the phase characteristics on the radar range, but the difference
between the expected and observed phase for each meteor trail. 
In contrast to \citep{Chisham_2013} the shape of the signal scattered by meteor trail 
is used for calibration, but not its average parameters produced by FitACF algorithm.

\begin{figure}
\includegraphics[scale=0.55]{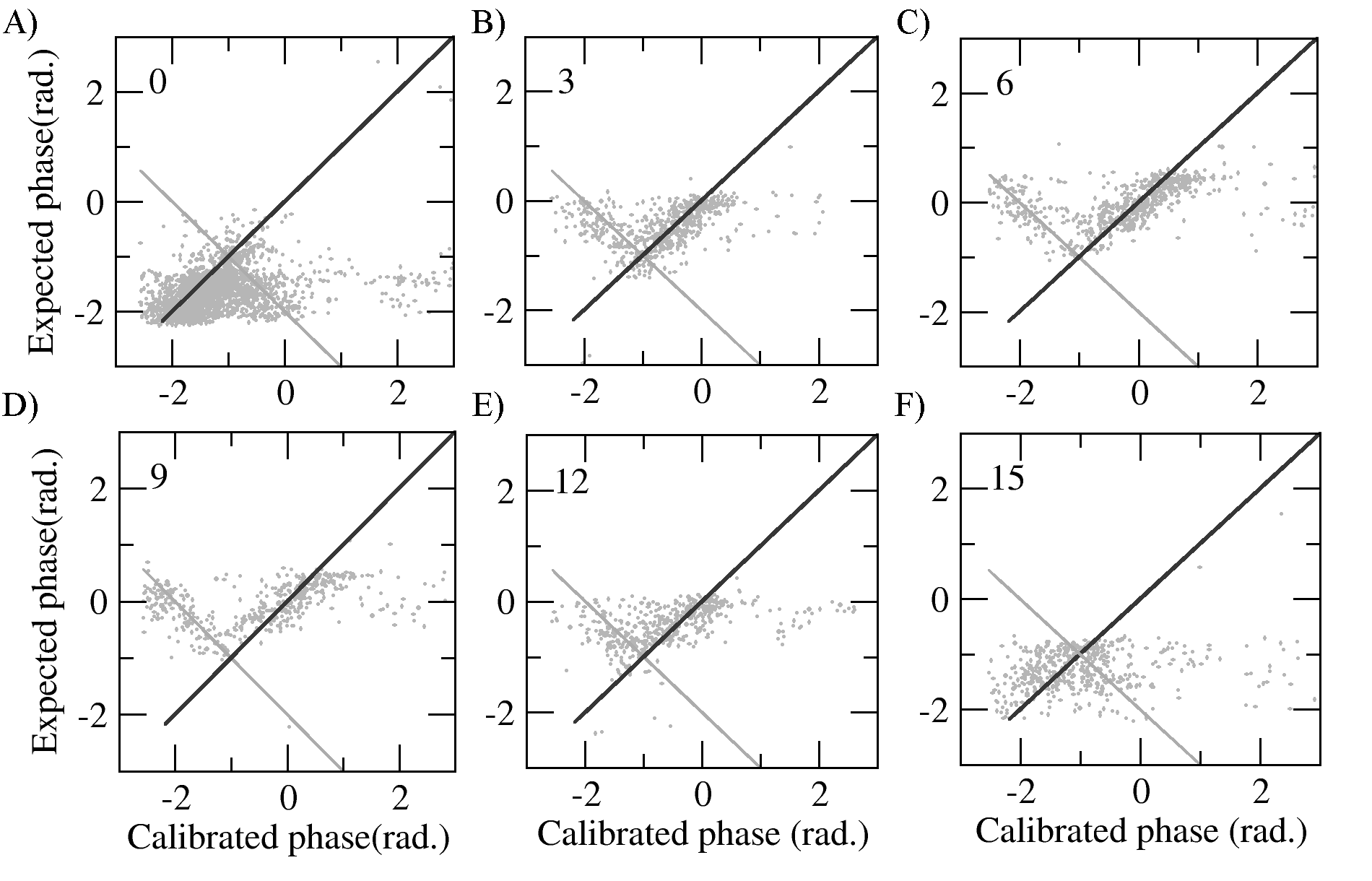}
\caption{Distribution of calibrated phases for signals 
scattered by meteor trails. Black line corresponds to 
the main lobe scattering, gray line - to the back lobe scattering.}
\label{fig:fig5}
\end{figure}

For the calibration, we processed about 2 months of interferometric
observations at the EKB radar - from 20/09/2019 to 18/11/2019. This results the 
calibration coefficient $A=-0.58$ radian. It differs slightly from
the first approximation $A_{0}=-0.49$ radian. This can be explained
by the small number of observations in the back lobe of the antenna
pattern. The resulting distributions of the expected phase and the
calibrated experimentally observed phase for different beams are shown
in Fig.\ref{fig:fig5}. It can be seen from the figure that the calibration
satisfactorily describes the experimental data on all the beams (the
expected linear dependence is shown by the black line). Variations of the observed
phase near the expected value can be associated both with a significant
dependence of the phase difference on the azimuth (caused by the hardware
used for the formation of the antenna
pattern) and with a significant level of the noise during the observations
- both auxiliary noise\citep{Berngardt_2018} and the noise of the analog
receivers.

\begin{figure}
\includegraphics[scale=0.25]{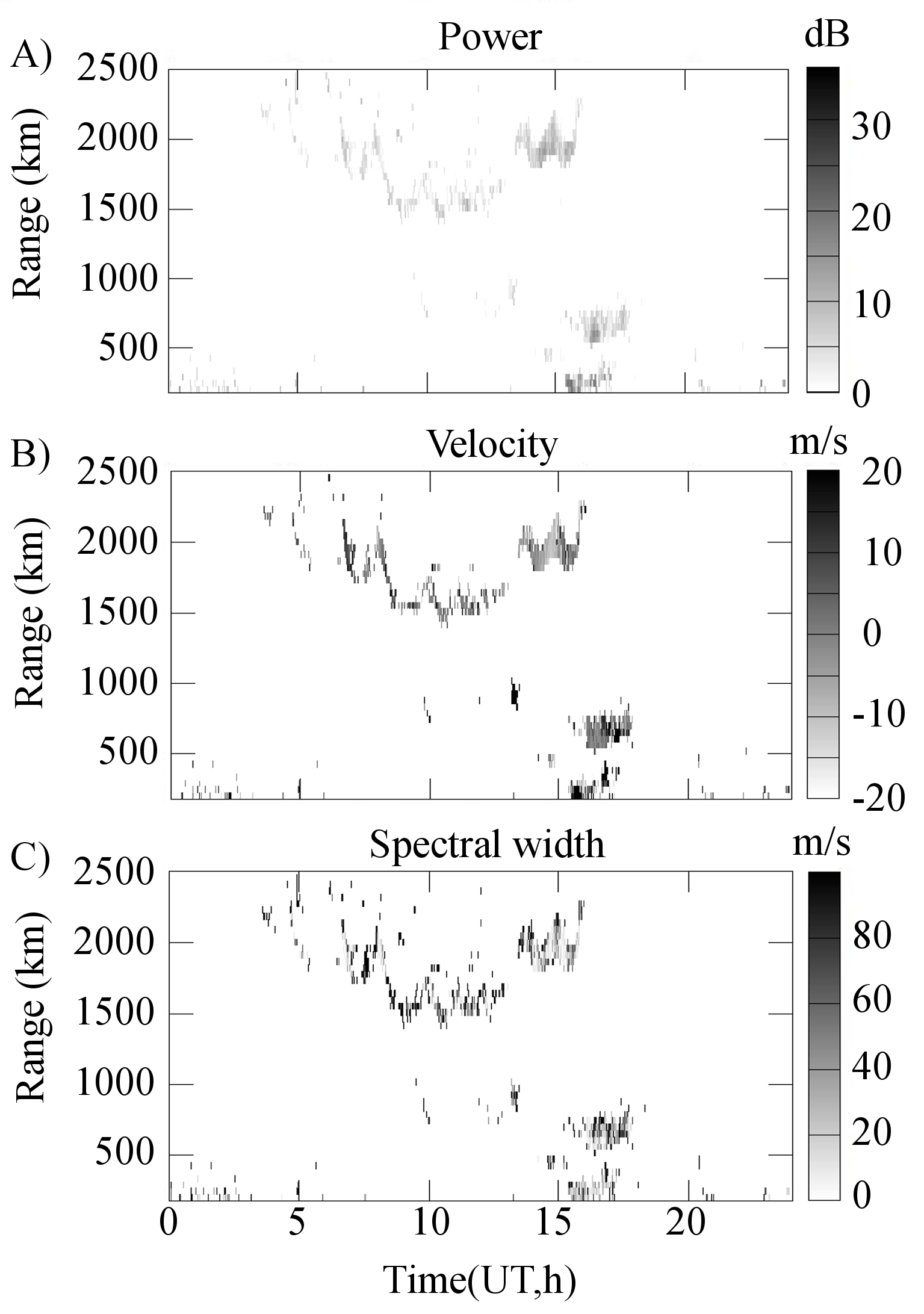}
\caption{Observed power(A), drift velocity(B) and spectral width(C) during 28/09/2019.}
\label{fig:fig6}
\end{figure}

The obtained results have been validated using technique \citep{Ponomarenko_2018} 
which estimates time delay $\delta T$ between main and interferometric channel. 
The calibation coefficient $A$ in this case becomes

\begin{equation}
A=  2\pi f_{0} \delta T
\end{equation}
This technique requires a statistically significant amount of E-layer echoes, 
so we applied to the data from 28/09/2019 when the E-layer scatter was 
clearly observed (see Fig.\ref{fig:fig6}). 
Our analysis produced the estimate of $\delta T = -8.4ns$.
For standard frequency used in this experiment 11.2MHz this 
corresponds to the calibration coefficient $A \approx -0.591$
in very good agreement with the technique presented in this paper.

\section{Testing the calibration results using FitACF data}

All the previous operations were carried out directly with the quadrature
components of the signal. However, when interpreting the experimental
data, a two-stage scheme for obtaining processed data is used: at
the first (RawACF) stage, the values of the autocorrelation and cross-correlation
functions are obtained (calculated respectively from the main array
signals and jointly from the main and interference arrays signals),
and at the second (FitACF) stage these correlation functions are used to estimate
the parameters of ionospheric irregularities using FitACF algorithm
\citep{Ribeiro_2013}. The calculation of the difference in interferometric
phases and the calculation of the
elevation angle is made at the second stage.

\begin{figure}
\includegraphics[scale=0.7]{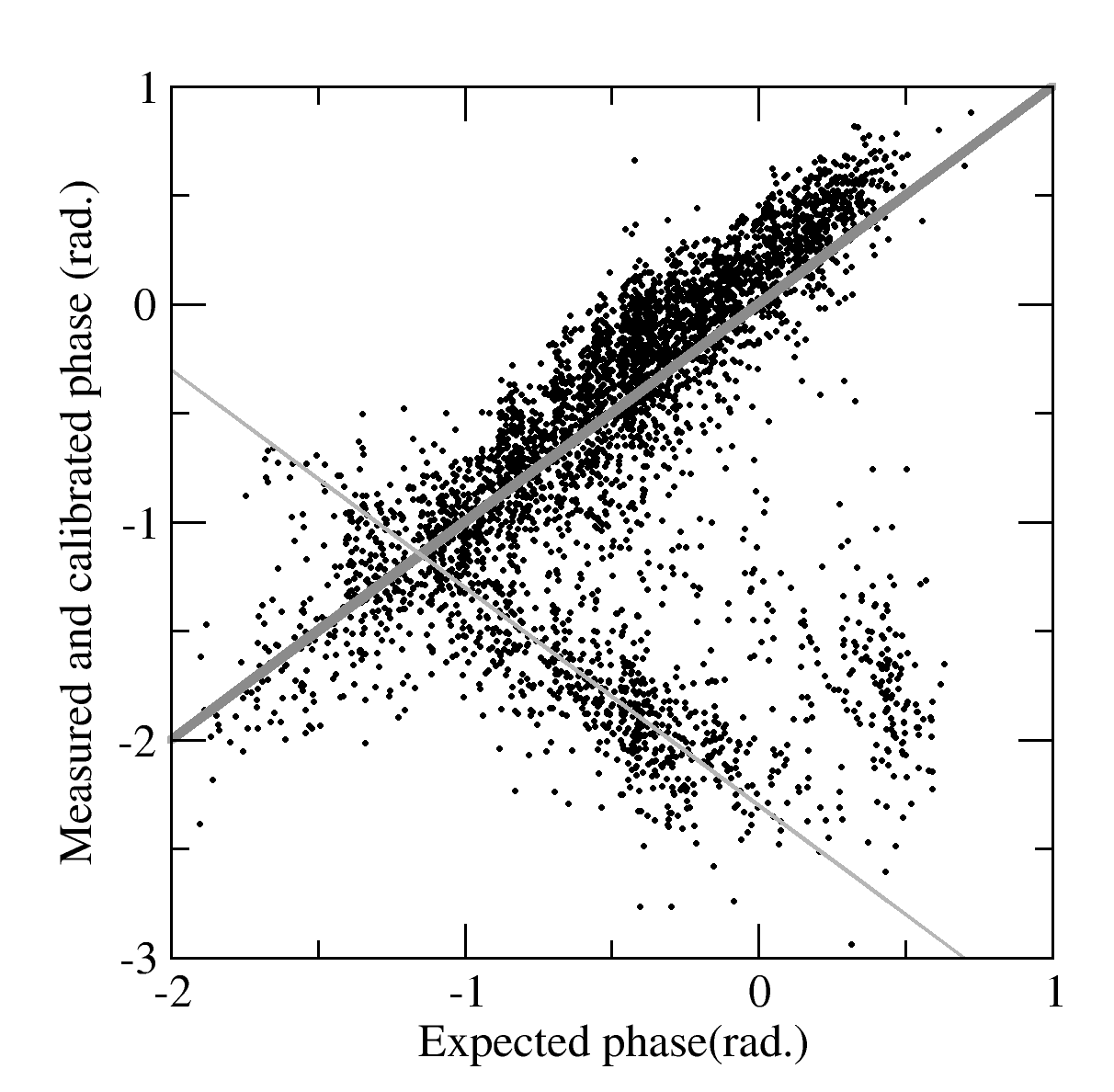}
\caption{Distribution of the expected phase for meteor trail scattering (calculated
from the measured range and standard height 90km) and the measured
calibrated phase for September-November 2019 obtained from processed
data (FitACF) over the ranges 250-750km. The thick line corresponds
to the scattering by meteors in the main lobe of the antenna pattern,
the thin line corresponds to the scattering in the back lobe of the
antenna pattern.}
\label{fig:fig7}
\end{figure}

Let us demonstrate the effectiveness of the calibration correction obtained
by this method for correcting the FitACF data. We studied
the measured interference phase obtained at the FitACF stage and calibrated
using calculated $A$. To do this we have processed the complete
set of interferometric observational data 20/09/2019 - 18/11/2019
at ranges \textless{}750km. For calibration we increase 
the interference phase from FitACF data by calibration coefficient
$A$ obtained in previous analysis and compare them with expected phase 
for meteor trail scattering. The comparison results are shown in Fig.\ref{fig:fig7}.
It can be seen from the figure that the received signals contain both
the signals scattered by the meteor trails from the main lobe of the
antenna pattern (thick line) and the signals from the back lobe (thin line). 
The figure
corresponds well to Fig.\ref{fig:fig4}D and Fig.\ref{fig:fig5} -
a few percent of the detected siqnals comes from the back lobe. 
In the bottom right corner of Fig.\ref{fig:fig7} there are also some signals 
of not meteoric nature - possibly E-layer scatter or near-range scatter.

\begin{figure}
\includegraphics[scale=0.6]{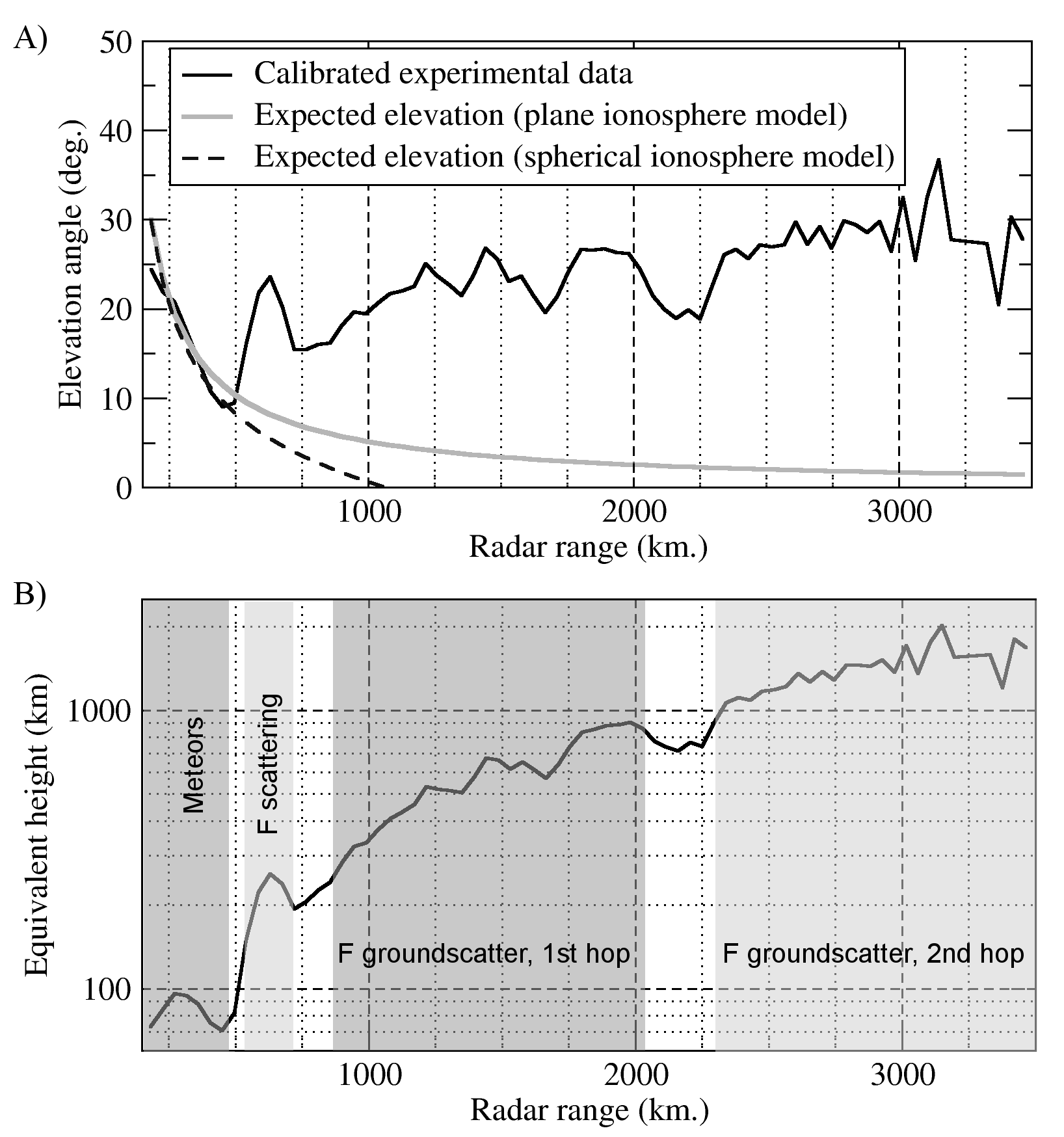}
\caption{Elevation angle statistics over September-November 2019 (black line), the expected 
elevation angle in the plane ionosphere model (gray solid line) and in the spherical 
ionospheric model (black dashed line) (A) and the
effective scattering height in the approximation of the linear radiowave
propagation calculated from these data (B) in the approximation of
a plane-layered ionosphere.}
\label{fig:fig8}
\end{figure}

To demonstrate the effectiveness of the method to calculate elevation
angles, we processed the same observational data during 20/09/2019-18/11/2019
at all the available radar ranges. The elevation angle was calculated
from the interference phase difference with taking into account the 
uncertainty of $2\pi n$, as a minimum value in
the range $0^{o}-45^{o}$ for $n\in[-5,5]$. A comparison between the calibrated
average elevation angle of the received signals and model elevation
angle for meteor observations calculated from geometric considerations
for a scattering height of 90km is shown in Fig.\ref{fig:fig8}A (in the two cases - 
for plane-layered ionosphere model and for spherical ionosphere model).
In the figure we used a plane-layered ionosphere for
a model elevation angle, that overestimates the
expected elevation angle, but gives a positive elevation angle for each range. Fig.\ref{fig:fig8}A shows a good agreement
between the average measured elevation angles and model ones at
the ranges less than 450 km, which confirms well the statement of
\citep{Chisham_2018} about predominant scattering from meteor trails
at ranges less than 400 km.

The ratio of the measured elevation angle $\alpha_{exp}$ to the model angle $\alpha_{mod}$ for 90
km scattering allows one to estimate the type of scattering under
no refraction assumption, using effective scattering height $h_{eff}$: 

\begin{equation}
h_{eff}=\frac{\alpha_{exp}}{\alpha_{mod}}\cdot 90km
\end{equation}

The mean effective scattering height is shown
in Fig.\ref{fig:fig8}. The effective heights above 300 km (taking into account
the errors of the plane-layered model) corresponds to a hop signal
propagation, therefore, in this experiment, at distances above 900
km the first hop ground scatter is most likely observed. Approximately
from the 2300km range (corresponding the effective heights above
900 km), a second hop ground scatter is most likely observed. The intermediate
regions most likely correspond to scattering by the ionospheric irregularities
of the E- and F- layers.

\begin{figure}
\includegraphics[scale=0.25]{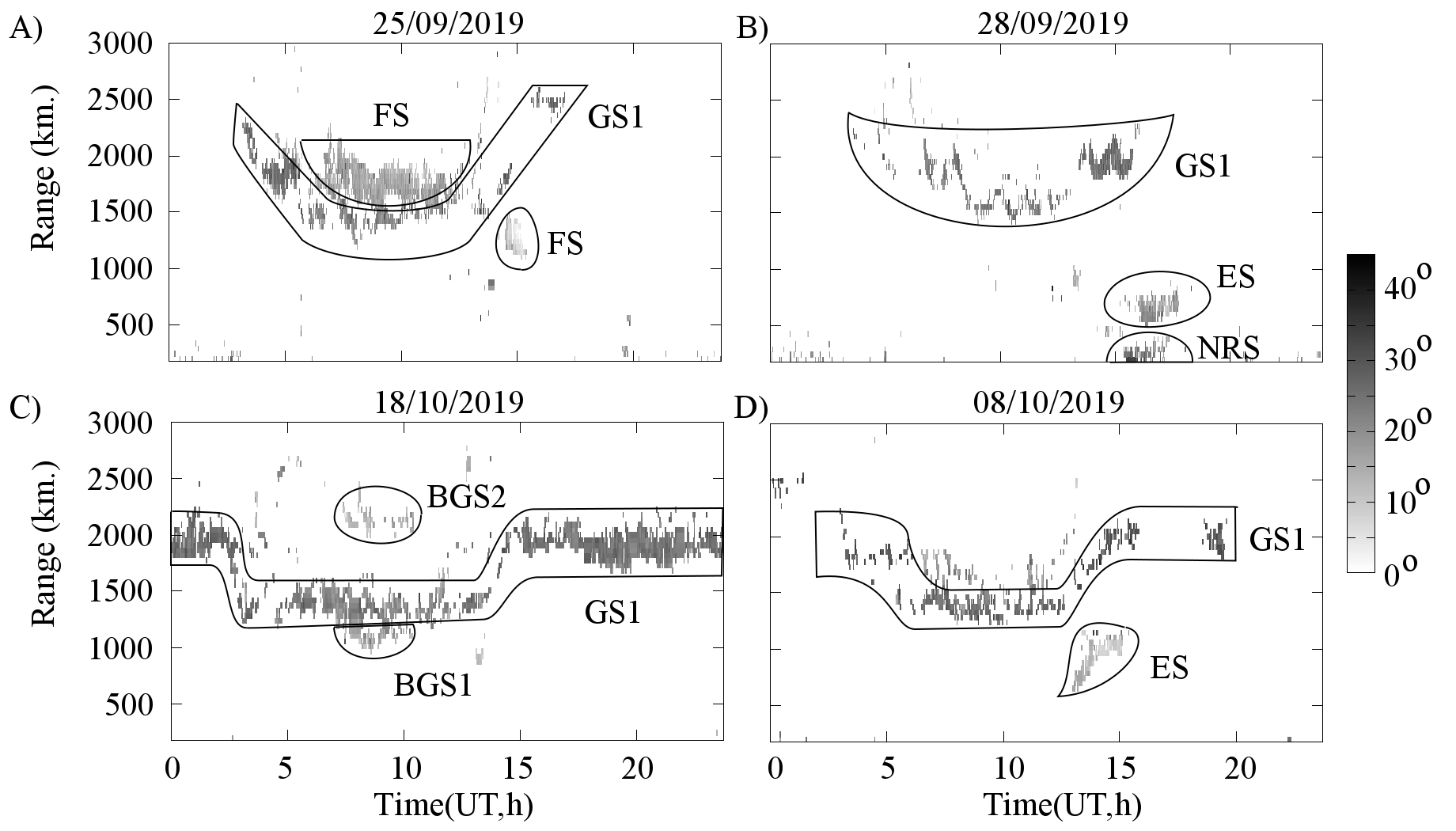}
\caption{Examples of calculations of elevation angles and approximate identification
of signal types in various experiments. GS1 is ground scatter, 1st hop;
BGS1 is ground scatter at back lobe, 1st hop; BGS2 is ground scatter at
back lobe, 2nd hop; FS is F region ionospheric scatter; ES is E-region
ionospheric scatter; NRS is near range scatter.}
\label{fig:fig9}
\end{figure}

Fig.\ref{fig:fig9} shows examples of calculated elevation 
in various experiments and an approximate identification of the 
scattered signal types depending on the elevation angle and range. Fig.\ref{fig:fig9}A
shows an example of ionospheric scatter in the F region moving from
high to low latitudes. Fig.\ref{fig:fig9}B shows an example of simultaneous
observation of ionospheric scatter in the E region and scattering
at near-range distances, related, judging by a very high elevation
angle (\textgreater{} 40 degrees), with possible reflections from
sporadic layers. Fig.\ref{fig:fig9}C illustrates the case of observing
ground scatter signals simultaneously in the main and back lobes of
the antenna pattern. Fig.\ref{fig:fig9}D shows an example of possible
scattering in the E-region.

\section{Conclusion}

The method for calibrating elevation measurements at EKB ISTP SB RAS
radar obtained for the period 20/09/2019 - 18/11/2019 is presented. 

The calibration method is a modernization of the method for calibrating
radar by meteor trails, proposed in \citep{Chisham_2013,Chisham_2018}.
The main difference of the method is the use of not a statistically
processed FitACF data, but the full waveform of the signals scattered on
the meteor trails. Using the full waveform makes it possible to more
reliably distinguish meteor scattering from other possible scattered
signal sources, and to determine meteor heights from the trail lifetime
using the NRLMSIS-00 model. Due
to the complex frequency dependence of the phase difference at the
EKB radar, the calibration was performed only for the frequency range
10-12 MHz, used for regular observations. A comparison of the results
with the method of \citep{Ponomarenko_2018} demonstrated a good agreement.

Based on a statistical analysis of the usability of the technique
was demonstrated on the processed FitACF data. It is shown that the obtained
elevation data on average corresponds well to the expected elevation
angles. It is shown that up to 450 km range, meteor scattering can
be considered dominant. The first examples of regular elevation observations
at the EKB ISTP SB RAS radar are presented, and their preliminary
interpretation is given.

\section*{Acknowledgments}
EKB ISTP SB RAS facility from Angara Center for Common Use of scientific equipment (http://ckp-rf.ru/ckp/3056/) is operated under budgetary funding of Basic Research program II.12. 
The data of EKB ISTP SB RAS radar are available at ISTP SB RAS (http://sdrus.iszf.irk.ru/ekb/page\_example/simple).
The work of OB,RF, and KG is supported by RFBR grant 18-05-00539a. 

\bibliography{REFS}

\begin{thebibliography}{32}
\expandafter\ifx\csname natexlab\endcsname\relax\def\natexlab#1{#1}\fi
\providecommand{\url}[1]{\texttt{#1}}
\providecommand{\href}[2]{#2}
\providecommand{\path}[1]{#1}
\providecommand{\DOIprefix}{doi:}
\providecommand{\ArXivprefix}{arXiv:}
\providecommand{\URLprefix}{URL: }
\providecommand{\Pubmedprefix}{pmid:}
\providecommand{\doi}[1]{\href{http://dx.doi.org/#1}{\path{#1}}}
\providecommand{\Pubmed}[1]{\href{pmid:#1}{\path{#1}}}
\providecommand{\bibinfo}[2]{#2}
\ifx\xfnm\relax \def\xfnm[#1]{\unskip,\space#1}\fi
\bibitem[{Berngardt et~al.(2018)Berngardt, Ruohoniemi, Nishitani, Shepherd,
  Bristow and Miller}]{Berngardt_2018}
\bibinfo{author}{Berngardt, O.}, \bibinfo{author}{Ruohoniemi, J.},
  \bibinfo{author}{Nishitani, N.}, \bibinfo{author}{Shepherd, S.},
  \bibinfo{author}{Bristow, W.}, \bibinfo{author}{Miller, E.},
  \bibinfo{year}{2018}.
\newblock \bibinfo{title}{{Attenuation of decameter wavelength sky noise during
  x-ray solar flares in 2013 - 2017 based on the observations of midlatitude HF
  radars}}.
\newblock \bibinfo{journal}{Journal of Atmospheric and Solar-Terrestrial
  Physics} \bibinfo{volume}{173}, \bibinfo{pages}{1 -- 13}.
\newblock \DOIprefix\doi{https://doi.org/10.1016/j.jastp.2018.03.022}.
\bibitem[{Berngardt et~al.(2015a)Berngardt, Kutelev, Kurkin, Grkovich,
  Yampolsky, Kashcheyev, Kashcheyev, Galushko, Grigorieva and
  Kusonsky}]{Berngardt_2015b}
\bibinfo{author}{Berngardt, O.I.}, \bibinfo{author}{Kutelev, K.A.},
  \bibinfo{author}{Kurkin, V.I.}, \bibinfo{author}{Grkovich, K.V.},
  \bibinfo{author}{Yampolsky, Y.M.}, \bibinfo{author}{Kashcheyev, A.S.},
  \bibinfo{author}{Kashcheyev, S.B.}, \bibinfo{author}{Galushko, V.G.},
  \bibinfo{author}{Grigorieva, S.A.}, \bibinfo{author}{Kusonsky, O.A.},
  \bibinfo{year}{2015}a.
\newblock \bibinfo{title}{{Bistatic Sounding of High-Latitude Ionospheric
  Irregularities Using a Decameter EKB Radar and an UTR-2 Radio Telescope:
  First Results}}.
\newblock \bibinfo{journal}{Radiophysics and Quantum Electronics}
  \bibinfo{volume}{58}, \bibinfo{pages}{390--408}.
\newblock \DOIprefix\doi{10.1007/s11141-015-9614-1}.
\bibitem[{Berngardt et~al.(2015b)Berngardt, Zolotukhina and
  Oinats}]{Berngardt_2015}
\bibinfo{author}{Berngardt, O.I.}, \bibinfo{author}{Zolotukhina, N.A.},
  \bibinfo{author}{Oinats, A.V.}, \bibinfo{year}{2015}b.
\newblock \bibinfo{title}{{Observations of field--aligned ionospheric
  irregularities during quiet and disturbed conditions with EKB radar: first
  results}}.
\newblock \bibinfo{journal}{Earth, Planets and Space} \bibinfo{volume}{67},
  \bibinfo{pages}{143}.
\newblock \DOIprefix\doi{10.1186/s40623-015-0302-3}.
\bibitem[{Bland et~al.(2014)Bland, McDonald, de~Larquier and
  Devlin}]{Bland_2014b}
\bibinfo{author}{Bland, E.C.}, \bibinfo{author}{McDonald, A.J.},
  \bibinfo{author}{de~Larquier, S.}, \bibinfo{author}{Devlin, J.C.},
  \bibinfo{year}{2014}.
\newblock \bibinfo{title}{{Determination of ionospheric parameters in real time
  using SuperDARN HF Radars}}.
\newblock \bibinfo{journal}{Journal of Geophysical Research: Space Physics}
  \bibinfo{volume}{119}, \bibinfo{pages}{5830--5846}.
\newblock \DOIprefix\doi{10.1002/2014JA020076}.
\bibitem[{Burrell et~al.(2016)Burrell, Yeoman, Milan and Lester}]{Burrell_2016}
\bibinfo{author}{Burrell, A.G.}, \bibinfo{author}{Yeoman, T.K.},
  \bibinfo{author}{Milan, S.E.}, \bibinfo{author}{Lester, M.},
  \bibinfo{year}{2016}.
\newblock \bibinfo{title}{Phase calibration of interferometer arrays at
  high-frequency radars}.
\newblock \bibinfo{journal}{Radio Science} \bibinfo{volume}{51},
  \bibinfo{pages}{1445--1456}.
\newblock \DOIprefix\doi{10.1002/2016rs006089}.
\bibitem[{Chisham(2018)}]{Chisham_2018}
\bibinfo{author}{Chisham, G.}, \bibinfo{year}{2018}.
\newblock \bibinfo{title}{{Calibrating SuperDARN Interferometers Using Meteor
  Backscatter}}.
\newblock \bibinfo{journal}{Radio Science} \bibinfo{volume}{53},
  \bibinfo{pages}{761--774}.
\newblock \DOIprefix\doi{10.1029/2017RS006492}.
\bibitem[{Chisham and Freeman(2013)}]{Chisham_2013}
\bibinfo{author}{Chisham, G.}, \bibinfo{author}{Freeman, M.},
  \bibinfo{year}{2013}.
\newblock \bibinfo{title}{{A reassessment of SuperDARN meteor echoes from the
  upper mesosphere and lower thermosphere}}.
\newblock \bibinfo{journal}{Journal of Atmospheric and Solar-Terrestrial
  Physics} , \bibinfo{pages}{--}\DOIprefix\doi{10.1016/j.jastp.2013.05.018}.
\bibitem[{Chisham et~al.(2007)Chisham, Lester, Milan, Freeman, Bristow,
  McWilliams, Ruohoniemi, Yeoman, Dyson, Greenwald, Kikuchi, Pinnock, Rash,
  Sato, Sofko, Villain and Walker}]{Chisham_2007}
\bibinfo{author}{Chisham, G.}, \bibinfo{author}{Lester, M.},
  \bibinfo{author}{Milan, S.}, \bibinfo{author}{Freeman, M.},
  \bibinfo{author}{Bristow, W.}, \bibinfo{author}{McWilliams, K.},
  \bibinfo{author}{Ruohoniemi, J.}, \bibinfo{author}{Yeoman, T.},
  \bibinfo{author}{Dyson, P.}, \bibinfo{author}{Greenwald, R.},
  \bibinfo{author}{Kikuchi, T.}, \bibinfo{author}{Pinnock, M.},
  \bibinfo{author}{Rash, J.}, \bibinfo{author}{Sato, N.},
  \bibinfo{author}{Sofko, G.}, \bibinfo{author}{Villain, J.P.},
  \bibinfo{author}{Walker, A.}, \bibinfo{year}{2007}.
\newblock \bibinfo{title}{{A decade of the Super Dual Auroral Radar Network
  (SuperDARN): scientific achievements, new techniques and future directions}}.
\newblock \bibinfo{journal}{Surveys in Geophysics} ,
  \bibinfo{pages}{33--109}\DOIprefix\doi{10.1007/s10712-007-9017-8}.
\bibitem[{Gillies et~al.(2011)Gillies, Hussey, Sofko, Ponomarenko and
  McWilliams}]{Gillies_2011}
\bibinfo{author}{Gillies, R.}, \bibinfo{author}{Hussey, G.},
  \bibinfo{author}{Sofko, G.}, \bibinfo{author}{Ponomarenko, P.},
  \bibinfo{author}{McWilliams, K.}, \bibinfo{year}{2011}.
\newblock \bibinfo{title}{{Improvement of HF coherent radar line-of-sight
  velocities by estimating the refractive index in the scattering volume using
  radar frequency shifting}}.
\newblock \bibinfo{journal}{Journal of Geophysical Research}
  \bibinfo{volume}{116}, \bibinfo{pages}{1302}.
\bibitem[{Greenwald et~al.(1995)Greenwald, Baker, Dudeney, Pinnock, Jones,
  Thomas, Villain, Cerisier, Senior, Hanuise, Hunsucker, Sofko, Koehler,
  Nielsen, Pellinen, Walker, Sato and Yamagishi}]{Greenwald_1995}
\bibinfo{author}{Greenwald, R.}, \bibinfo{author}{Baker, K.B.},
  \bibinfo{author}{Dudeney, J.R.}, \bibinfo{author}{Pinnock, M.},
  \bibinfo{author}{Jones, T.}, \bibinfo{author}{Thomas, E.},
  \bibinfo{author}{Villain, J.P.}, \bibinfo{author}{Cerisier, J.C.},
  \bibinfo{author}{Senior, C.}, \bibinfo{author}{Hanuise, C.},
  \bibinfo{author}{Hunsucker, R.D.}, \bibinfo{author}{Sofko, G.},
  \bibinfo{author}{Koehler, J.}, \bibinfo{author}{Nielsen, E.},
  \bibinfo{author}{Pellinen, R.}, \bibinfo{author}{Walker, A.},
  \bibinfo{author}{Sato, N.}, \bibinfo{author}{Yamagishi, H.},
  \bibinfo{year}{1995}.
\newblock \bibinfo{title}{{Darn/Superdarn: A Global View of the Dynamics of
  High--Lattitude Convection}}.
\newblock \bibinfo{journal}{Space Science Reviews} \bibinfo{volume}{71},
  \bibinfo{pages}{761--796}.
\newblock \DOIprefix\doi{10.1007/BF00751350}.
\bibitem[{Greenwood et~al.(2011)Greenwood, Parkinson, Dyson and
  Schulz}]{Greenwood_2011}
\bibinfo{author}{Greenwood, R.}, \bibinfo{author}{Parkinson, M.},
  \bibinfo{author}{Dyson, P.}, \bibinfo{author}{Schulz, E.},
  \bibinfo{year}{2011}.
\newblock \bibinfo{title}{{Dominant ocean wave direction measurements using the
  TIGER SuperDARN systems}} \bibinfo{volume}{73}, \bibinfo{pages}{2379--2385}.
\bibitem[{Holdsworth et~al.(2004)Holdsworth, Reid and Cervera}]{7770504}
\bibinfo{author}{Holdsworth, D.A.}, \bibinfo{author}{Reid, I.M.},
  \bibinfo{author}{Cervera, M.A.}, \bibinfo{year}{2004}.
\newblock \bibinfo{title}{{Buckland Park all-sky interferometric meteor
  radar}}.
\newblock \bibinfo{journal}{Radio Science} \bibinfo{volume}{39},
  \bibinfo{pages}{1--12}.
\newblock \DOIprefix\doi{10.1029/2003rs003014}.
\bibitem[{Hosokawa et~al.(2005)Hosokawa, Ogawa, Arnold, Lester, Sato and
  Yukimatu}]{Hosokawa_2005}
\bibinfo{author}{Hosokawa, K.}, \bibinfo{author}{Ogawa, T.},
  \bibinfo{author}{Arnold, N.F.}, \bibinfo{author}{Lester, M.},
  \bibinfo{author}{Sato, N.}, \bibinfo{author}{Yukimatu, A.},
  \bibinfo{year}{2005}.
\newblock \bibinfo{title}{Extraction of polar mesosphere summer echoes from
  superdarn data}.
\newblock \bibinfo{journal}{Geophys. Res. Lett.} \bibinfo{volume}{32},
  \bibinfo{pages}{L12801}.
\newblock \DOIprefix\doi{10.1029/2005gl022788}.
\bibitem[{Hosokawa et~al.(2004)Hosokawa, Ogawa, Yukimatu, Sato and
  Iyemori}]{Hosokawa_2004}
\bibinfo{author}{Hosokawa, K.}, \bibinfo{author}{Ogawa, T.},
  \bibinfo{author}{Yukimatu, A.}, \bibinfo{author}{Sato, N.},
  \bibinfo{author}{Iyemori, T.}, \bibinfo{year}{2004}.
\newblock \bibinfo{title}{Statistics of antarctic mesospheric echoes observed
  with the superdarn syowa radar}.
\newblock \bibinfo{journal}{Geophys. Res. Lett.} \bibinfo{volume}{31},
  \bibinfo{pages}{L02106}.
\newblock \DOIprefix\doi{10.1029/2003gl018776}.
\bibitem[{{Jones} and {Jones}(1990)}]{Jones_and_Jones1990JATP...52..185J}
\bibinfo{author}{{Jones}, W.}, \bibinfo{author}{{Jones}, J.},
  \bibinfo{year}{1990}.
\newblock \bibinfo{title}{{Ionic diffusion in meteor trains}}.
\newblock \bibinfo{journal}{Journal of Atmospheric and Terrestrial Physics}
  \bibinfo{volume}{52}, \bibinfo{pages}{185--191}.
\newblock \DOIprefix\doi{10.1016/0021-9169(90)90122-4}.
\bibitem[{Koustov et~al.(2007)Koustov, Andre, Turunen, Raito and
  Milan}]{Koustov_2007}
\bibinfo{author}{Koustov, A.}, \bibinfo{author}{Andre, D.},
  \bibinfo{author}{Turunen, E.}, \bibinfo{author}{Raito, T.},
  \bibinfo{author}{Milan, S.}, \bibinfo{year}{2007}.
\newblock \bibinfo{title}{{Heights of SuperDARN F region echoes estimated from
  the analysis of HF radio wave propagation}} \bibinfo{volume}{25},
  \bibinfo{pages}{1987--1994}.
\bibitem[{Lester et~al.(2004)Lester, Chapman, Cowley, Crooks, Davies, Hamadyk,
  McWilliams, Milan, Parsons, Payne, Thomas, Thornhill, Wade, Yeoman and
  Barnes}]{Lester_2004}
\bibinfo{author}{Lester, M.}, \bibinfo{author}{Chapman, P.},
  \bibinfo{author}{Cowley, S.}, \bibinfo{author}{Crooks, S.},
  \bibinfo{author}{Davies, J.}, \bibinfo{author}{Hamadyk, P.},
  \bibinfo{author}{McWilliams, K.}, \bibinfo{author}{Milan, S.},
  \bibinfo{author}{Parsons, M.}, \bibinfo{author}{Payne, D.},
  \bibinfo{author}{Thomas, E.}, \bibinfo{author}{Thornhill, J.},
  \bibinfo{author}{Wade, N.}, \bibinfo{author}{Yeoman, T.},
  \bibinfo{author}{Barnes, R.}, \bibinfo{year}{2004}.
\newblock \bibinfo{title}{{Stereo CUTLASS - A new capability for the SuperDARN
  HF radars}} \bibinfo{volume}{22}, \bibinfo{pages}{459--473}.
\bibitem[{{MMANA-GAL software}()}]{MMANAsite}
\bibinfo{author}{{MMANA-GAL software}}, .
\newblock \bibinfo{title}{{MMANA-GAL software}}.
\newblock \URLprefix \url{http://gal-ana.de/basicmm/en/}.
\bibitem[{Nishitani et~al.(2019)Nishitani, Ruohoniemi, Lester, Baker, Koustov,
  Shepherd, Chisham, Hori, Thomas, Makarevich, Marchaudon, Ponomarenko, Wild,
  Milan, Bristow, Devlin, Miller, Greenwald, Ogawa and
  Kikuchi}]{Nishitani_2019}
\bibinfo{author}{Nishitani, N.}, \bibinfo{author}{Ruohoniemi, J.},
  \bibinfo{author}{Lester, M.}, \bibinfo{author}{Baker, J.B.H.},
  \bibinfo{author}{Koustov, A.V.}, \bibinfo{author}{Shepherd, S.G.},
  \bibinfo{author}{Chisham, G.}, \bibinfo{author}{Hori, T.},
  \bibinfo{author}{Thomas, E.G.}, \bibinfo{author}{Makarevich, R.A.},
  \bibinfo{author}{Marchaudon, A.}, \bibinfo{author}{Ponomarenko, P.},
  \bibinfo{author}{Wild, J.A.}, \bibinfo{author}{Milan, S.E.},
  \bibinfo{author}{Bristow, W.A.}, \bibinfo{author}{Devlin, J.},
  \bibinfo{author}{Miller, E.}, \bibinfo{author}{Greenwald, R.A.},
  \bibinfo{author}{Ogawa, T.}, \bibinfo{author}{Kikuchi, T.},
  \bibinfo{year}{2019}.
\newblock \bibinfo{title}{{Review of the accomplishments of mid-latitude Super
  Dual Auroral Radar Network (SuperDARN) HF radars}}.
\newblock \bibinfo{journal}{Progress in Earth and Planetary Science}
  \bibinfo{volume}{6}, \bibinfo{pages}{27}.
\newblock \DOIprefix\doi{10.1186/s40645-019-0270-5}.
\bibitem[{{NRLMSISE-00}()}]{NRLMSISsite}
\bibinfo{author}{{NRLMSISE-00}}, .
\newblock \bibinfo{title}{{NRLMSISE-00 site}}.
\newblock \URLprefix
  \url{https://ccmc.gsfc.nasa.gov/pub/modelweb/atmospheric/msis/nrlmsise00/}.
\bibitem[{Ogunjobi et~al.(2015)Ogunjobi, Sivakumar, Stephenson and
  Sivla}]{Ogunjobi_2015}
\bibinfo{author}{Ogunjobi, O.}, \bibinfo{author}{Sivakumar, V.},
  \bibinfo{author}{Stephenson, J.A.E.}, \bibinfo{author}{Sivla, W.T.},
  \bibinfo{year}{2015}.
\newblock \bibinfo{title}{{Evidence of Polar Mesosphere Summer Echoes Observed
  by SuperDARN SANAE HF Radar in Antarctica}}.
\newblock \bibinfo{journal}{Terrestrial, Atmospheric and Oceanic Sciences}
  \bibinfo{volume}{26}, \bibinfo{pages}{431}.
\newblock \DOIprefix\doi{10.3319/tao.2015.03.06.01(aa)}.
\bibitem[{Picone et~al.(2002)Picone, Hedin, Drob and Aikin}]{NRLMSIS_2002}
\bibinfo{author}{Picone, J.M.}, \bibinfo{author}{Hedin, A.E.},
  \bibinfo{author}{Drob, D.P.}, \bibinfo{author}{Aikin, A.C.},
  \bibinfo{year}{2002}.
\newblock \bibinfo{title}{{NRLMSISE-00 empirical model of the atmosphere:
  Statistical comparisons and scientific issues}}.
\newblock \bibinfo{journal}{Journal of Geophysical Research: Space Physics}
  \bibinfo{volume}{107}, \bibinfo{pages}{SIA 15--1--SIA 15--16}.
\newblock \DOIprefix\doi{10.1029/2002JA009430}.
\bibitem[{Ponomarenko et~al.(2016)Ponomarenko, Iserhienrhien and
  St.-Maurice}]{Ponomarenko_2016}
\bibinfo{author}{Ponomarenko, P.}, \bibinfo{author}{Iserhienrhien, B.},
  \bibinfo{author}{St.-Maurice, J.P.}, \bibinfo{year}{2016}.
\newblock \bibinfo{title}{{Morphology and possible origins of near-range
  oblique HF backscatter at high and midlatitudes}}.
\newblock \bibinfo{journal}{Radio Science} \bibinfo{volume}{51},
  \bibinfo{pages}{718--730}.
\newblock \DOIprefix\doi{10.1002/2016rs006088}.
\bibitem[{Ponomarenko et~al.(2015)Ponomarenko, Nishitani, Oinats, Tsuya and
  St.-Maurice}]{Ponomarenko_2015}
\bibinfo{author}{Ponomarenko, P.}, \bibinfo{author}{Nishitani, N.},
  \bibinfo{author}{Oinats, A.V.}, \bibinfo{author}{Tsuya, T.},
  \bibinfo{author}{St.-Maurice, J.P.}, \bibinfo{year}{2015}.
\newblock \bibinfo{title}{{Application of ground scatter returns for
  calibration of HF interferometry data}}.
\newblock \bibinfo{journal}{Earth, Planets and Space} \bibinfo{volume}{67},
  \bibinfo{pages}{138}.
\newblock \DOIprefix\doi{10.1186/s40623-015-0310-3}.
\bibitem[{Ponomarenko et~al.(2010)Ponomarenko, St-Maurice, Hussey and
  Koustov}]{Ponomarenko_2010}
\bibinfo{author}{Ponomarenko, P.}, \bibinfo{author}{St-Maurice, J.P.},
  \bibinfo{author}{Hussey, G.}, \bibinfo{author}{Koustov, A.},
  \bibinfo{year}{2010}.
\newblock \bibinfo{title}{{HF ground scatter from the polar cap: Ionospheric
  propagation and ground surface effects}}.
\newblock \bibinfo{journal}{J. Geophys. Res} \bibinfo{volume}{115},
  \bibinfo{pages}{10310}.
\newblock \DOIprefix\doi{10.1029/2010JA015828}.
\bibitem[{Ponomarenko et~al.(2018)Ponomarenko, St.-Maurice and
  McWilliams}]{Ponomarenko_2018}
\bibinfo{author}{Ponomarenko, P.}, \bibinfo{author}{St.-Maurice, J.P.},
  \bibinfo{author}{McWilliams, K.}, \bibinfo{year}{2018}.
\newblock \bibinfo{title}{{Calibrating HF Radar Elevation Angle Measurements
  Using E Layer Backscatter Echoes}}.
\newblock \bibinfo{journal}{Radio Science} \bibinfo{volume}{53},
  \bibinfo{pages}{1438--1449}.
\newblock \DOIprefix\doi{10.1029/2018rs006638}.
\bibitem[{Ponomarenko et~al.(2009)Ponomarenko, St-Maurice, Waters, Gillies and
  Koustov}]{Ponomarenko_2009}
\bibinfo{author}{Ponomarenko, P.}, \bibinfo{author}{St-Maurice, J.P.},
  \bibinfo{author}{Waters, C.L.}, \bibinfo{author}{Gillies, R.},
  \bibinfo{author}{Koustov, A.}, \bibinfo{year}{2009}.
\newblock \bibinfo{title}{{Refractive index effects on the scatter volume
  location and Doppler velocity estimates of ionospheric HF backscatter
  echoes}}.
\newblock \bibinfo{journal}{Ann. Geophys.} \bibinfo{volume}{27},
  \bibinfo{pages}{4207--4219}.
\newblock \DOIprefix\doi{10.5194/angeo-27-4207-2009}.
\bibitem[{Ribeiro et~al.(2013)Ribeiro, Ruohoniemi, Ponomarenko, N.~Clausen,
  Baker, Greenwald, Oksavik and de~Larquier}]{Ribeiro_2013}
\bibinfo{author}{Ribeiro, A.J.}, \bibinfo{author}{Ruohoniemi, J.},
  \bibinfo{author}{Ponomarenko, P.}, \bibinfo{author}{N.~Clausen, L.B.},
  \bibinfo{author}{Baker, J.}, \bibinfo{author}{Greenwald, R.},
  \bibinfo{author}{Oksavik, K.}, \bibinfo{author}{de~Larquier, S.},
  \bibinfo{year}{2013}.
\newblock \bibinfo{title}{{A comparison of SuperDARN ACF fitting methods}}.
\newblock \bibinfo{journal}{Radio Science} \bibinfo{volume}{48},
  \bibinfo{pages}{274--282}.
\newblock \DOIprefix\doi{10.1002/rds.20031}.
\bibitem[{Tsutsumi et~al.(1999)Tsutsumi, Holdsworth, Nakamura and
  Reid}]{Tsutsumi_BPMR_autodetect1999}
\bibinfo{author}{Tsutsumi, M.}, \bibinfo{author}{Holdsworth, D.},
  \bibinfo{author}{Nakamura, T.}, \bibinfo{author}{Reid, I.},
  \bibinfo{year}{1999}.
\newblock \bibinfo{title}{Meteor observations with an mf radar}.
\newblock \bibinfo{journal}{Earth, Planets and Space} \bibinfo{volume}{51},
  \bibinfo{pages}{691--699}.
\newblock \URLprefix \url{https://doi.org/10.1186/BF03353227},
  \DOIprefix\doi{10.1186/BF03353227}.
\bibitem[{Tsutsumi et~al.(2009)Tsutsumi, Yukimatu, Holdsworth and
  Lester}]{Tsutsumi_SDARN_meteorwindIQ}
\bibinfo{author}{Tsutsumi, M.}, \bibinfo{author}{Yukimatu, A.},
  \bibinfo{author}{Holdsworth, D.A.}, \bibinfo{author}{Lester, M.},
  \bibinfo{year}{2009}.
\newblock \bibinfo{title}{{Advanced SuperDARN meteor wind observations based on
  raw time series analysis technique}}.
\newblock \bibinfo{journal}{Radio Sci.} \bibinfo{volume}{44},
  \bibinfo{pages}{RS2006}.
\newblock \DOIprefix\doi{10.1029/2008rs003994}.
\bibitem[{Villain et~al.(1984)Villain, Greenwald and Vickrey}]{Villain_1984}
\bibinfo{author}{Villain, J.P.}, \bibinfo{author}{Greenwald, R.A.},
  \bibinfo{author}{Vickrey, J.F.}, \bibinfo{year}{1984}.
\newblock \bibinfo{title}{{HF ray tracing at high latitudes using measured
  meridional electron density distributions}}.
\newblock \bibinfo{journal}{Radio Science} \bibinfo{volume}{19},
  \bibinfo{pages}{359--374}.
\newblock \URLprefix
  \url{https://agupubs.onlinelibrary.wiley.com/doi/abs/10.1029/RS019i001p00359},
  \DOIprefix\doi{10.1029/RS019i001p00359}.
\bibitem[{Yukimatu and Tsutsumi(2002)}]{Yukimatu_2002}
\bibinfo{author}{Yukimatu, A.S.}, \bibinfo{author}{Tsutsumi, M.},
  \bibinfo{year}{2002}.
\newblock \bibinfo{title}{{A new SuperDARN meteor wind measurement: Raw time
  series analysis method and its application to mesopause region dynamics}}.
\newblock \bibinfo{journal}{Geophysical Research Letters} \bibinfo{volume}{29},
  \bibinfo{pages}{42--1--42--4}.
\newblock \DOIprefix\doi{10.1029/2002GL015210}.

\end{thebibliography}
\end{document}